\documentstyle[aps,preprint,epsf]{revtex}
\newcommand{\beq}{\begin{equation}}
\newcommand{\eeq}{\end{equation}}
\begin{document}
\draft
\tightenlines

\title{On the relation between the Hartree-Fock and  Kohn-Sham approaches}

\author{M.Ya. Amusia$^{a,b}$,
A.Z. Msezane$^{c}$, V.R. Shaginyan$^{c,d}$,
\footnote{E--mail: vrshag@thd.pnpi.spb.ru}
D. Sokolovski$^{e}$}
\address{$^{a}$Racah Institute of Physics,
the Hebrew University, Jerusalem 91904, Israel;
$^{b}$A.F. Ioffe Physical-Technical Institute, 194021 St.
Petersburg, Russia;\\
$^{c}$CTSPS, Clark Atlanta University, Atlanta,
Georgia 30314, USA;\\
$^{d}$Petersburg Nuclear Physics Institute,
Gatchina, 188300, Russia\\
$^{e}$Queen's University of Belfast, Belfast BT7 1NN, UK}
\maketitle

\begin{abstract}

We show that the Hartree-Fock (HF) results cannot be reproduced
within the framework of Kohn-Sham (KS) theory because the
single-particle densities of finite systems obtained within
the HF calculations are not $v$-representable,
i.e., do not correspond to
any ground state of a $N$ non-interacting electron systems
in a local external potential.
For this reason, the KS theory, which finds
a minimum on a different subset of all densities,
can overestimate the ground state energy, as compared to
the HF result.
The discrepancy between the two approaches provides
no grounds to assume that either the KS theory or the
density functional theory suffers from
internal contradictions.

\end{abstract}


The Hartree-Fock method (HF), first proposed in the pioneering  works of
Hartree and Fock \cite{har,foc} is known to be successful in
calculating properties of electron systems, in particular,
the ground state properties of atoms. Based on
a variational principle, the HF method estimates the
ground state energy $E$ of an electron system from
above, i.e., $E_{HF}\ge E$, where $E_{HF}$ is the ground state energy
calculated within the HF method. If the ground state wave function
of $N$ electrons is approximated by a single $N$-electron Slater
determinant, the HF solution delivers the minimum value $E_{HF}$ on the
set of all such determinants.  Agreement, or otherwise, with the HF
results is often used to estimate the success of other approximate
computational schemes.

The Density Functional Theory (DFT)
exploits the one-to-one correspondence
between the single-particle electron density and an external
potential acting upon the system and relies on
the existence of a universal functional $F[\rho({\bf r})]$
which can be minimized in order to find the ground state energy
\cite{hwk}.
The Kohn-Sham (KS) theory goes further in reducing the problem of
calculating ground state properties of a many-electron system in a
local external single-particle potential to solving Hartree-like
one-electron equations \cite{hwk,wks}.
Successful solution of these equations allows to
predict, at least in principle, the atomic, molecular, cluster and
solid bodies binding energies, phonon spectra, activation barriers
etc., see e.g. \cite {parr}.

It is natural, therefore, to ask whether the HF ground state
energy can be successfully reproduced in the Kohn-Sham
approach.
We note first that
a universal density functional $F_{HF}[\rho]$ can be
{\it defined}
with the help of the  constrained-search technique \cite{levy}.
Had the explicit form of $F_{HF}[\rho]$ been available,
the HF and the DFT approach would have yielded the same
results for the ground state energy, $E_{HF}=E_{DHF}$, and the electron
density, $\rho_{HF}({\bf r})=\rho_{DHF}({\bf r})$, \cite{levy}.
Here, $E_{DHF}$ and $\rho_{DHF}({\bf r})$ are the exact DFT
Hartree-Fock energy and density respectively, while
$\rho_{HF}({\bf r})$ is the HF density.
Unfortunately, the correspondence theorem \cite{hwk,levy}
establishes the existence of the functional only in
principle,
and provides no unique practical recipe for its construction.
Rather, for practical calculations one has to resort
to the KS approach.
Exhaustive calculations  \cite{goer,nes,ns1}
of the ground state energies of different
atoms show that, if the KS approach is used,
the resulting energy $E_{KSHF}$  usually exceeds
the energy $E_{HF}$ obtained by the HF method,
\begin{equation}
E_{KSHF}\geq E_{HF}.
\end{equation}
Recently, there have been suggestions that this disagreement
may point to intrinsic flaws in both the DFT and the KS
theories. One might suspect, therefore,
that an exact local exchange potential does not exist
for ground states of typical atoms (see \cite{nes,ns1} and
references therein).
We will, however, argue that
the discrepancy (1) between the HF and KS
is due to the different domains on which the respective
functionals are defined. More specifically,
we will show that while a KS density is $v$-representative,
a HF density is not, i.e., it cannot
be obtained as the ground state density of any
$N$ non-interacting electrons in a local potential.
As a result, the KS method simply delivers a minimum
on a different class of electron densities,
and its disagreement with the HF approach does not
indicate the existence of any internal contradictions
either in KS or DFT approach.

The purpose of this Letter is to
show that HF results cannot be reproduced
within the framework of KS theory because the
single-particle densities of finite
$N$ electron systems obtained within the HF calculations are not
$v$-representable.
This means that the HF densities cannot
be obtained as the ground state density of any
$N$ non-interacting electrons in a local potential.
Thus, the discrepancy between the two approaches, which manifests
itself in the inequality (1), provides
no grounds to assume that either the KS theory or the
DFT suffers from internal contradictions.

We begin our study with  considering the HF ground state energy
which is given by the equation
\begin{eqnarray}
E_{HF} & = & -\frac{1}{2}\sum\limits_{j=1}^{N} n_j\int\phi_j^{\ast}
({\bf r})\nabla^{2}\phi_j({\bf r})d{\bf r}
+\frac{1}{2}\sum\limits_{j=1,i=1}^{N} n_j n_i
\int\frac{\phi_j^{\ast}({\bf r}_{1})
\phi_i^{\ast}({\bf r}_{2})\phi_j({\bf r}_{1})
\phi_i({\bf r}_{2})}{|{\bf r}_{1}-{\bf r}_{2}|}d{\bf r}_{1}d{\bf
r}_{2}\\
& + & E_{x}
+\sum\limits_{j=1} n_j\int\phi_j^{\ast}({\bf r})\phi_j({\bf
r})v({\bf r})d{\bf r}.\nonumber  \end{eqnarray}
We use an atomic system of units: $e=m=\hbar =1$, where $e$ and $m$ are electron
charge and mass, respectively.
Here $N$ is the total number of electrons,
$n_i$ are the occupation numbers:
$n_i=1$ if the corresponding
single-particle level is occupied, otherwise $n_i=0$.
For an atom one has
$v({\bf r})=-Z/r$, where $Z$ is the nuclear charge.
The exchange energy $E_x$ can be represented as follows
\begin{eqnarray}
E_{x}& = & -
\frac{1}{2}\int
\frac{\chi_{0}({\bf r}_{1},{\bf r}_{2},iw)+2\pi
\rho({\bf r}_{1})\delta(w)\delta({\bf r}_{1}-{\bf r}_{2})}{|{\bf
r}_{1}-{\bf r}_{2}|}\frac{dw}{2\pi}d{\bf r}_{1}d{\bf r}_{2}\\
& = & -\sum_{k,i}n_{k}n_{i}\int\left[\frac{\phi_{i}^{\ast }({\bf
r}_{1})\phi_{i}({\bf r}_{2})\phi_{k}^{\ast }({\bf r}_{2})\phi
_{k}({\bf r}_{1})}{|{\bf r}_{1}-{\bf r}_{2}|}\right] d{\bf
r}_{1}d{\bf r}_{2} \nonumber,
\end{eqnarray} and $\chi_{0}({\bf r}_{1},{\bf r}_{2},\omega)$
is the linear response
function, which is of the form \begin{equation} \chi_{0}({\bf
r}_{1},{\bf r}_{2},\omega )=\sum_{i,k}n_{i}(1-n_{k})\phi_{i}^{\ast
}({\bf r}_{1})\phi_{i}({\bf r}_{2})\phi_{k}^{\ast }({\bf r}
_{2})\phi_{k}({\bf r}_{1})\left[\frac{1}{\omega -\omega_{ik}+i\eta }-
\frac{1}{\omega +\omega_{ik}-i\eta }\right],
\end{equation}
with $\omega_{ik}$ defined as $\omega_{ik}=\varepsilon
_{k}-\varepsilon_{i}$, $\varepsilon_{k}$ and functions $\phi
_{k}({\bf r}_{1})$ being respectively the one-particle energies and
wave functions of the HF equations; and $\eta$ is
the infinitely small number, $\eta\rightarrow 0$.
Varying  Eq. (2) with respect to the
single-particle wave functions $\phi_{i}({\bf r})$, one obtains the
HF system of
equations:  \beq
\left[-\frac{\nabla^{2}}{2}+v({\bf r})\right]
\phi_{i}^{HF}({\bf r})\\
+ \sum\limits_{j=1}^{N}\int
\frac{d{\bf r}^{\prime}}
{|{\bf r}-{\bf r}^{\prime }|}
\left[|\phi_{j}^{HF}({\bf r}^{\prime })|^{2}
\phi_{i}^{HF}({\bf r})-\phi_{j}^{HF\ast }({\bf r}^{\prime })
\phi_{i}^{HF}({\bf r}^{\prime
})\phi_{j}^{HF}({\bf r})\right]=
E_{i}^{HF}\phi_{i}^{HF}({\bf r})
\eeq
These equations differ from an ordinary Schr\"{o}dinger equation in
two essential aspects: they are non-linear in
$\phi_{i}^{HF}({\bf r})$ and the second
term under the sum on the left hand side of Eq. (5)
that represents the so-called Fock's potential is non-local.

The asymptotic behavior in $r$ of $\phi_{i}^{HF}({\bf r})$, contrary to
the case of an ordinary one-particle Schr\"{o}dinger
equation, is not determined by $E_{i}^{HF}$ and does not have the form
\begin{equation}
\phi_{i}({\bf r})|_{r\rightarrow\infty }\sim\exp
(-\sqrt{2|E_{i}^{HF}|}r).  \end{equation}
On the contrary, it was shown in \cite{hms} that at $r\rightarrow
\infty $ the function $\phi_{i}^{HF}({\bf r})$ is not determined by
$E_{i}^{HF}$ but behaves as \begin{equation}\phi_{i}^{HF}({\bf
r})|_{r\rightarrow\infty }\sim\sum\nolimits_{l}C_{l}[\exp
(-\sqrt{2|E_{F}^{HF}|}r)]/r^{l+1}, \end{equation} where $E_{F}^{HF}$
is the energy of the so-called Fermi-level (with wave function $\phi
_{F}^{HF}({\bf r})$), which is the smallest binding energy
of the occupied single-particle levels among all
$E_{i}$ in the considered system, and \begin{equation}
C_{l}=\sum\limits_{m=-l}^{+l}\int
\phi_{F}^{HF\ast }({\bf r})r^{l}
{\bf Y}_{lm}({\bf r}/r)\phi_{i}^{HF}({\bf r})d{\bf r},
\end{equation}
and ${\bf Y}_{lm}({\bf r}/r)$ is the $l^{th}$ order spherical
polynomial. The uniform behavior of the occupied levels given by Eq. (7) is a
consequence of the long range nature of the non-local Fock potential.
The uniform behavior (7) leads to
a number of very specific features of the ground state
wave function, which, as we will see, cannot occur in the KS equations
and makes it impossible for the KS equations (see Eq. (13)) to reproduce
the HF density.
That is, it makes the HF densities be non-$v$-representable.

Consider next the HF calculations based on the constrained-search
formulation of DFT which employs the HF density functional
$F_{HF}[\rho]$ \cite{levy}. The functional
$F_{HF}[\rho]$ is obtained by minimizing the expectation
value $F_{HF}[\rho]=(\Psi\lbrack\rho]|\hat{H}|\Psi\lbrack\rho])$
over all single-electron Slater
determinants consistent with a fixed density $\rho({\bf r})$.
The Hamiltonian $\hat{H}$ is of the form
\begin{equation} \hat{H}=-\frac{1}{2}\int\hat{\psi}^{\ast}
({\bf r})\nabla^{2}\hat{\psi}({\bf r})d{\bf r}
+\frac{1}{2}\int\frac{\hat{\psi}^{\ast}({\bf r}_{1})\hat{
\psi}^{\ast}({\bf r}_{2})\hat{\psi}({\bf r}_{2})\hat{\psi}({\bf
r}_{1})}{|{\bf r}_{1}-{\bf r}_{2}|}d{\bf r}_{1}d{\bf r}_{2}.
\end{equation}
From the manner in which the functional
$F_{HF}[\rho]$ is constructed it is clear that the
resulting ground state energy
$E_{DHF}$ equals that obtained in the HF method,
$E_{HF}$,
\beq E_{DHF}=E_{HF}=F_{HF}[\rho_{DHF}]
+\int v({\bf r})\rho_{DHF}({\bf r})d{\bf r}.
\eeq
Obviously, $\rho_{DHF}({\bf r})=\rho_{HF}({\bf r})$,
where $\rho_{HF}({\bf r})$ is the HF density obtained upon
solving Eq. (5), while the determinant $\Psi^{HF}$,
which yields the minimum value, is
composed of the eigenfunctions of Eq. (5) \cite{levy}.

To give proof of the non-$v$ representability of the HF densities,
we {\it assume for a moment} that $\rho_{HF}({\bf r})$ is
non-interacting $v$-representable i.e., that it can be
represented as the ground state density of $N$ non-interacting electrons
described by the Schr\"{o}dinger equation
with some local potential $v_L({\bf r})$. Equivalently,
we {\it assume} that the functional
$F_{HF}[\rho]$ is defined in the domain of $v$-representable
densities. As a result, the HF functional
$F_{HF}[\rho]$ can be represented as $F_{HF}[\rho]\equiv F_{x}[\rho]$
with $F_x[\rho]$ being a functional defined in the domain of
$v$-representable densities,
\beq
F_{x}[\rho]=T_k[\rho]
+\frac{1}{2}\sum\limits_{j=1,i=1}^{N} n_j n_i
\int\frac{\phi_j^{\ast}({\bf r}_{1})
\phi_i^{\ast}({\bf r}_{2})\phi_j({\bf r}_{1})
\phi_i({\bf r}_{2})}{|{\bf r}_{1}-{\bf r}_{2}|}
d{\bf r}_{1}d{\bf
r}_{2}+E_x
+\int v({\bf r})\rho({\bf r})d{\bf r}.\eeq
The second term in Eq. (11) being the Hartree term is obviously a
functional of the density.
The KS kinetic energy functional $T_k[\rho]$
is known to be defined in the domain of $v$-representable densities
\cite{wks,goer} and so is the fourth terms in Eq. (11).
Thus, the exchange energy $E_x$  given by Eq. (3)
is also to be a functional $E_x[\rho]$, defined on the $v$-representable
densities, as has been
shown in \cite{s2}, with the help of the representation (3).
It has also been demonstrated in \cite{s2} that
the variational derivative of $E_x[\rho]$ exists and can be
evaluated explicitly to produce the KS exchange potential
\cite{s2} \begin{equation} V_{x}({\bf r})=\frac{\delta E_{x}[\rho]}
{\delta \rho({\bf r})}.\end{equation}
We can then proceed to obtain the eigenvalues
$\varepsilon_i$ and the wave functions
$\phi_i({\bf r})$ in Eqs. (4) and (11),
by solving the KS single-particle
equations
\begin{equation} \left(
-\frac{\nabla^{2}}{2}
+ \sum\limits_{j=1}^{N}\int
\frac{d{\bf r}^{\prime}}
{|{\bf r}-{\bf r}^{\prime }|}
|\phi_{j}({\bf r}^{\prime })|^{2}
+V_{x}({\bf r})+v({\bf r})\right)\phi_{i}({\bf r})
=\varepsilon_{i}\phi_{i}({\bf r}).
\end{equation}
and compute the density $\rho({\bf r})$ as
\begin{equation} \rho({\bf
r})=\sum_{i}n_{i}|\phi_{i}({\bf r})|^{2}.\end{equation}
We note that due to the  constrained-search technique \cite{levy}
the minimum of the functional $F_x[\rho]$ is given by
\beq
F_{x}[\rho]=(\Phi_{KS}|\hat{H}|\Phi_{KS})
+\int v({\bf r})\rho({\bf r})d{\bf r},
\eeq
where $\Phi_{KS}$ is a single $N$-electron Slater determinant which
delivers the lowest energy expectation value of $\hat{H}$ given by Eq.
(9). This determinant is composed of the single-particle wave
functions $\phi_{i}({\bf r})$ which are the solutions of one-particle
equations (13). Thus, we are led to the conclusion that the
determinant $\Psi^{HF}$ has to coincide with the determinant
$\Psi_{KS}$. Then the HF wave functions $\phi_i^{HF}$
and the eigenvalues $E^{HF}_i$ given by Eq. (5) must
be equal to the wave-functions $\phi_i$ and
the eigenvalues $\varepsilon_i$
given by Eq. (13). However, it is  seen from Eq. (7) that
all single-particle HF functions have the same asymptotic behavior
determined by the smallest orbital energy \cite{hms},
even though the eigenvalues $E^{HF}_i$ are not, in general,
degenerate. On the other hand, eigenfunctions of Eq. (13) may exhibit
such behavior only if the eigenvalues $\varepsilon_i$ are
degenerate as it follows from Eq. (6). As a result,
we arrived at  contradiction.
This contradiction is resolved once we
recognize that at least some of the HF densities are not
non-interacting $v$-representable.
Note that examples of
non-$v$-representable densities
were given in Ref. \cite{levy1}. Obviously, a
one-to-one correspondence between non-local potentials and local
ones does not exist \cite{tlg}. Therefore, if the explicit
form the functionals $F^{HF}[\rho]$ and $T_k[\rho]$ was known,
we would not have been
able to obtain the HF ground state within the KS theory
because they are defined on the different subsets of densities.
For example, it is impossible to reproduce the HF single-particle
eigenvalues $E_{i}^{HF}$ within the KS theory \cite{ams}.  In other
words, the two approximate methods have different domains of
applicability and are not amenable to a direct comparison.
Consequently, the result (1) cannot be
used to prove that the KS method or DFT is in any way deficient,
as was suggested in Refs.\cite{nes,ns1}.

A few remarks are in order here.
A single $N$-electron Slater determinant coinciding with
$\Phi_{KS}$ can be obtained in
the optimized effective potential method
with the local exchange potential $V_{OPM}$ \cite{tsh}.
Because of one-to-one correspondence that exists
between the wave function, the density  and the
local single-particle potential \cite{hwk,levy},
$V_{OPM}$ must coincide with $V_{x}$ given by Eq. (12),
$V_{x}({\bf r})=V_{OPM}({\bf r})$
as has been demonstrated in
\cite{s2}. Therefore, the ground state energy $E_{OEP}$ of a
many-electron system calculated in OPM has to be equal to the
corresponding energy $E_{KSHF}$ calculated
with the functional $F_x[\rho]$, $E_{KSHF}=E_{OEP}$.
It is now tempting to assume that all three energies agree, i.e., that
$E_{HF}=E_{KSHF}=E_{OEP}$ \cite{nes,ns1}.
But it is not the case since the HF densities are not
$v$-representable. This fact is in agreement with
numerical calculations showing that
$E_{HF}<E_{KSHF}$ \cite{goer,nes,ns1}.
One should also mention one special case where
the HF and KS theories give the same answer.
For a He atom, the HF potential acting on the occupied
states is local and the HF density is $v$-representable,
so that, $E_{OEP}=E_{KSHF}=E_{DHF}=E_{HF}$.
This observation is confirmed by numerical calculations \cite{nes}. For
all other atoms, the HF potentials are non-local, and
the HF densities are not $v$-representable.
Obviously, in that case,
one has $E_{OEP}=E_{KSHF}>E_{DHF}=E_{HF}$ in accordance with
the numerical calculations \cite{nes}.

In summary, by clarifying the
relationship between the non-local exchange HF potential and the
local exchange KS potential, we have shown
that the Hartree-Fock method cannot be reproduced
within the framework of Kohn-Sham theory because the
single-particle densities of finite systems obtained in
Hartree-Fock calculations are not $v$-representable.
We have demonstrated that
the fact that the KS calculations of finite electron systems
lead to higher ground
state energies
cannot be used to infer the existence of inconsistencies
in either KS or DFT theory.
Most of the specific features of the HF method
result from the non-local nature of the  HF potential.
For this reason, they provide no grounds to criticize
the Kohn-Sham theory which deals with local single-particle
potentials and $v$-representable densities.
To conclude, it is worth mentioning that at present there is no
compelling evidence to believe the HF method to be superior
to the KS approach.
Among the drawbacks of the HF theory is
the well-known fact that
the HF non-local single-particle potential acting on the unoccupied states
falls off exponentially. As a result, the HF potential can only support
very few unoccupied energy levels,
which leads to difficulties in treating the excited states. By
contrast, the KS theory does not suffer from the drawbacks
inherent in  the
HF method. Thus, a further study is needed to clarify
which type of behavior actually occurs in atoms
with a large number of electrons and whether
the failure to agree with the HF results can, indeed,
be considered a fault of the KS theory.

The visit of VRS and DS to Clark Atlanta University has been supported by
NSF through a grant to CTSPS. MYaA is grateful to the S.A.
Shonbrunn Research Fund for support of his research.
AZM is supported by US DOE, Division of Chemical Sciences, Office of
Basic Energy Sciences, Office of Energy Research.
This work was supported in part by INTAS, project no. 03-51-6170.

\end{document}